\documentclass{article}
\usepackage{amsmath,amssymb}
\usepackage[margin=3cm]{geometry}
\usepackage{hyperref} 
\newcommand{\til}{\tilde}

\def\m@th{\mathsurround=0pt}
\mathchardef\bracell="0365 %% the number identifies the character, in this case: tilde
\def\upbrall{$\m@th\bracell$}
\def\undertilde#1{\mathop{\vtop{\ialign{##\crcr
    $\hfil\displaystyle{#1}\hfil$\crcr
     \noalign
     {\kern1.5pt\nointerlineskip} %% moves the character further down
     \upbrall\crcr\noalign{\kern1pt %% puts more space underneath
   }}}}\limits}

\mathchardef\hatbracell="0362 %% the number identifies the character, in this case: hat
\def\hatupbrall{$\m@th\hatbracell$}
\def\underhat#1{\mathop{\vtop{\ialign{##\crcr
    $\hfil\displaystyle{#1}\hfil$\crcr
     \noalign
     {\kern1.5pt\nointerlineskip} %% moves the character further down
     \hatupbrall\crcr\noalign{\kern1pt %% puts more space underneath
   }}}}\limits}

\newcommand{\cL}{\mathcal{L}}
\newcommand{\Li}{{\rm Li}}
\newcommand{\Dl}{\Delta}
\newcommand{\dl}{\delta}
\newcommand{\sg}{\sigma}
\newcommand{\lm}{\lambda}

\newcommand{\bv}{\boldsymbol{v}}
\newcommand{\bze}{{\boldsymbol 0}}
\newcommand{\ptl}{\partial}

\newcommand{\nn}{\nonumber}

\newcommand{\be}{\begin{equation}}
\newcommand{\ee}{\end{equation}}
\newcommand{\bea}{\begin{eqnarray}}
\newcommand{\eea}{\end{eqnarray}}
\newcommand{\bse}{\begin{subequations}}
\newcommand{\ese}{\end{subequations}}

\begin{document}

\title{An integrable multicomponent quad equation and its Lagrangian formulation}
\author{J Atkinson, SB Lobb and FW Nijhoff}
\maketitle

\abstract{We present a hierarchy of discrete systems whose first members are the lattice modified Korteweg-de Vries equation, and the lattice modified Boussinesq equation. The $N$-th member in the hierarchy is an $N$-component system defined on an elementary plaquette in the 2-dimensional lattice.
The system is multidimensionally consistent and a Lagrangian which respects this feature, i.e., which has the desirable closure property, is obtained.}

\section{Introduction}
What we now call the lattice modified Korteweg-de Vries (MKdV) equation appeared, up to a point-transformation, as early as 1894 in the work of Bianchi \cite{Bia1894} as the permutability condition for B\"acklund transformations of the sine-Gordon equation. It appeared again as an integrable partial difference equation via Hirota's method in \cite{Hir1977}, and it was derived via the direct linearization scheme in \cite{NijQuiCap1983}.

To describe the lattice MKdV in the modern setting, let $v=v(l,m)$ be the dependent variable depending on position $(l,m)$ in a 2-dimensional lattice,  where $l$ and $m$ shift by integer values, and let $p,q$ be parameters associated with the $l,m$-lattice directions respectively. Then the lattice MKdV equation reads
\be\label{MKdV}
 p(v\hat{v}-\til{v}\hat{\til{v}}) = q(v\til{v}-\hat{v}\hat{\til{v}}),
\ee
where ``$\;\til{}\;$'' denotes a shift in the $l$-direction, and ``$\;\hat{}\;$'' denotes a shift in the $m$-direction.

A generalized direct linearization scheme proposed in \cite{NijPapCapQui1992} led to a higher order analogue of the lattice MKdV equation, which can be identified as a lattice version of the modified Boussinesq (MBSQ) equation of \cite{QuiNijCap1982}. This is a scalar equation on a stencil of 9 lattice points, given by
\be\label{scalarMBSQ} \biggl(\frac{p^2\hat{\til{v}}-q^2\hat{\hat{v}}}{p\hat{\hat{v}}-q\hat{\til{v}}}\biggr)\frac{\hat{\hat{\til{v}}}}{\hat{v}}-\biggl(\frac{p^2\til{\til{v}}-q^2\hat{\til{v}}}{p\hat{\til{v}}-q\til{\til{v}}}\biggr)\frac{\hat{\til{\til{v}}}}{\til{v}}
 = p\biggl(\frac{v}{\til{v}}-\frac{\hat{\hat{\til{v}}}}{\hat{\hat{\til{\til{v}}}}}\biggr)-q\biggl(\frac{v}{\hat{v}}-\frac{\hat{\til{\til{v}}}}{\hat{\hat{\til{\til{v}}}}}\biggr).
\ee
A more convenient form of this equation appeared in \cite{Nij1999}, specifically a coupled system of equations in two dependent variables $v,w$ defined on an elementary plaquette: 
\be\label{MBSQ}
 \frac{p\hat{v}-q\til{v}}{\hat{\til{v}}} = \frac{p\til{v}\hat{w}-q\hat{v}\til{w}}{v\hat{\til{w}}} = \frac{p\til{w}-q\hat{w}}{w}.
\ee
The scalar equation \eqref{scalarMBSQ} can be derived from \eqref{MBSQ} by eliminating one or other of the variables. As shown in \cite{Nij1999}, 
in either form this is a discrete analogue of the potential modified Boussinesq equation \cite{QuiNijCap1982}.  
Symmetries and conservation laws for the two-component form of the equation were derived in \cite{XenNij2012}.  

The lattice Gel'fand-Dikii (GD) hierarchy, that is the system given in \cite{NijPapCapQui1992} which goes to the continuous GD hierarchy under suitable limits\footnote{In the 1970s members of the Russian school, including Gel'fand and Dikii, studied higher order spectral problems generalizing aspects of the Korteweg-de Vries equation (c.f. e.g. \cite{GelDik1976ii,Man1979}); hence these types of systems have been referred to as Gel'fand-Dikii hierarchies.}, is a multicomponent system whose first members are the lattice potential KdV equation and a lattice version of the Boussinesq equation. 
A lattice `modified' GD hierarchy was also proposed in the Appendix of \cite{NijPapCapQui1992}, the first two members of which, (\ref{MKdV}) and (\ref{scalarMBSQ}), were given explicitly, but its higher members were left implicit due to the difficulty of writing corresponding scalar equations in explicit form on increasing stencil size.

In this paper we study a hierarchy of systems whose first two members are \eqref{MKdV} and \eqref{MBSQ}. Different to the direct linearization approach of \cite{NijPapCapQui1992} it will be obtained by a reduction procedure from the discrete Kadomtsev-Petviashvili (KP) equation of Hirota \cite{Hir1981}, and in fact the precise relationship between the system studied here and those of \cite{NijPapCapQui1992} is not yet known beyond 
the first two cases.
Like the first cases (\ref{MKdV}) and (\ref{MBSQ}), each system is defined on an elementary plaquette of the 2-dimensional lattice (i.e., it is a quad equation); in particular each is consistent in multidimensions and possesses a Lax representation. Our main objectives are to present the derivation of this hierarchy as well as a Lagrangian formulation that respects its consistency property.

The new kind of variational principle for multidimensionally consistent systems proposed in \cite{LobNij2009} relies on a closure property of the associated Lagrangian, allowing it to be interpreted as a closed form. A Lagrangian with this property was given in \cite{LobNij2010} for the generic member of the lattice GD hierarchy. We give here a Lagrangian for the generic member of the hierarchy in this paper, for which there is a clear connection to that of the discrete bilinear KP equation, whose Lagrangian was given in \cite{LobNijQui2009}.

In Section \ref{Section_system} we derive the $N$-component lattice system, give the Lax pair, describe its multidimensional consistency and give some important symmetries. In Section \ref{Section_Lagrangian} we present the Lagrangian structure, and in Section \ref{Section_closure} we demonstrate that this Lagrangian obeys a closure relation on solutions of the system. Section \ref{Section_conclusion} contains some additional remarks.

\section{The system}\label{Section_system}
It was observed in \cite{Atk2008} that the one-variable and two-variable equations (\ref{MKdV}) and (\ref{MBSQ}) can be obtained by a natural reduction procedure from the discrete KP equation of Hirota \cite{Hir1981}.
Generalization of this reduction procedure then led to a quadrilateral lattice equation in $N$ variables $v_1,\ldots,v_N$ that can be written as
\be\label{discreteMGD}
 \frac{p\hat{v}_{n-1}\til{v}_{n}-q\til{v}_{n-1}\hat{v}_{n}}{\hat{\til{v}}_{n-1}v_{n}} = \frac{p\hat{v}_{n}\til{v}_{n+1}-q\til{v}_{n}\hat{v}_{n+1}}{\hat{\til{v}}_{n}v_{n+1}}, \;\;\;\;\;\; n\in\{1,\dots,N\},
\ee
where by convention\footnote{This convention is a convenient notational device which avoids the need to separate equations such as (\ref{discreteMGD}) into sub cases $n=1$, $n\in\{2,\ldots,N-1\}$ and $n=N$.} $v_0=v_{N+1}=1$. 
This system, which contains (\ref{MKdV}) and (\ref{MBSQ}) as cases $N=1$ and $N=2$, is the main object studied in the present paper.

\subsection{Reduction from Hirota's discrete KP}
We begin by giving the reduction procedure from which (\ref{discreteMGD}) may be obtained.
The starting point is the Hirota discrete KP equation \cite{Hir1981}
\be\label{notHirotaMiwa}
 \bar{\tau}\hat{\til{\tau}} = p\hat{\tau}\til{\bar{\tau}}-q\til{\tau}\hat{\bar{\tau}},
\ee
where $\tau=\tau(l,m,n)$, $\til{\tau}=\tau(l+1,m,n)$, $\hat{\tau}=\tau(l,m+1,n)$, $\bar{\tau}=\tau(l,m,n+1)$ etc. are values of the dependent variable $\tau$ as a function of independent variables $l,m,n\in\mathbb{Z}$.
In this equation $p,q\in\mathbb{C}\setminus\{0\}$ are parameters, they are often taken equal to 1, which up to a non-autonomous change of variables is without loss of generality.
Such change of variables will however not be possible after the reduction procedure.
From the consistency property of (\ref{notHirotaMiwa}) there can be obtained a B\"acklund transformation as follows:
\bse
\label{hbt}
\bea
&&\bar{\tau}\til{\sg} = p\til{\bar{\tau}}\sg - \lambda\til{\tau}\bar{\sg},\\
&&\bar{\tau}\hat{\sg} = q\hat{\bar{\tau}}\sg - \lambda\hat{\tau}\bar{\sg}.
\eea
\ese
Here, if $\tau=\tau(l,m,n)$ is a solution of (\ref{notHirotaMiwa}) then the coupled system (\ref{hbt}) for $\sigma$ is consistent and determines a function $\sigma=\sigma(l,m,n)$ also satisfying (\ref{notHirotaMiwa}).
This is a weak B\"acklund transformation\footnote{This terminology was probably first used by McCarthy \cite{McC1978} to distinguish the situation where an equation is strictly sufficient for compatibility of the B\"acklund system, as opposed to a strong B\"acklund transformation, where an equation emerges as both necessary and sufficient.}, in fact the equation emerging as compatibility constraint for (\ref{hbt}) is 
\be\label{10pteqn}
 \frac{p\hat{\tau}\til{\bar{\tau}}-q\til{\tau}\hat{\bar{\tau}}}{\bar{\tau}\hat{\til{\tau}}} = \biggl[\frac{p\hat{\tau}\til{\bar{\tau}}-q\til{\tau}\hat{\bar{\tau}}}{\bar{\tau}\hat{\til{\tau}}}\biggr]^{-},
\ee
and it can be directly verified that (\ref{hbt}) is a B\"acklund transformation, in the strong sense, for this equation.
The equation \eqref{10pteqn} lies on a 10-point stencil on the three-dimensional lattice, by integration it can be seen as a non-autonomous version of (\ref{notHirotaMiwa}).

The equation (\ref{discreteMGD}) governs solutions of (\ref{10pteqn}) satisfying the following additional constraints
\bse
\label{constraints}
\bea
\label{constraint1}
&&\tau(l,m,n+N+1)=\tau(l,m,n), \quad l,m,n\in\mathbb{Z},\\
\label{constraint2}
&&\tau(l,m,0)=1, \quad l,m\in\mathbb{Z},
\eea
\ese
which can be seen by identifying
\be
v_{n}(l,m):=\tau(l,m,n), \quad l,m\in\mathbb{Z}, \quad n \in\{1,\ldots, N\}.
\ee
The first constraint (\ref{constraint1}) is simply an imposed periodicity, the second constraint (\ref{constraint2}) is not so usual, it consists of some particular boundary data.
\subsection{Integrability}
The original B\"acklund transformation (\ref{hbt}) of equation (\ref{10pteqn}) does not preserve the constraints (\ref{constraints}) but it is consistent with the first constraint (\ref{constraint1}).
That is, we may write 
\be
\sigma(l,m,n+N+1)=\sigma(l,m,n), \quad l,m,n\in\mathbb{Z}\label{constraint3}
\ee
without breaking the consistency.
The further observation that this B\"acklund transformation is already linear in $\sigma$ makes it, through this reduction procedure, a Lax pair for (\ref{discreteMGD}).
In other words defining
\be
\phi_{n}(l,m):=\sigma(l,m,n), \quad l,m\in\mathbb{Z}, \quad n \in\{0,\ldots, N\},
\ee
and imposing (\ref{constraints}) and (\ref{constraint3}) reduces (\ref{hbt}) to the system
\be \label{Lax}
\til{\Phi} = L_{p}(\bv,\til{\bv})\Phi,\quad
\hat{\Phi} = L_{q}(\bv,\hat{\bv})\Phi,
\ee
where \be\Phi := [\phi_0,\phi_1,\ldots,\phi_N]^T, \quad \bv:=(v_1,\ldots,v_N),\label{notation}\ee and
\be
 L_{p}(\bv,\til{\bv}) = \left(\begin{array}{cccccc}
                        p\til{v}_{1}/v_{1} & -\lm/v_{1}         & 0                     & \cdots             & \;\;\;\;\;\;\cdots & 0\\
                        0                  & p\til{v}_{2}/v_{2} & -\lm\til{v}_{1}/v_{2} & 0                  & \;\;\;\;\;\;\cdots & 0\\
                                           &                    &                       &                    &                    & \vdots\\
                        \vdots             &                    & \;\;\;\;\;\;\ddots    & \;\;\;\;\;\;\ddots & \;\;\;\;\;\;\ddots & \vdots\\
                                           &                    &                       &                    &                    & 0\\
                        0                  & \cdots             & \cdots                & 0                  & p\til{v}_{N}/v_{N} & -\lm\til{v}_{N-1}/v_{N}\\
                        -\lm\til{v}_{N}    & 0                  & \cdots                &                    & 0                  & p
                        \end{array}\right)
\ee
is the $(N+1)$-square Lax matrix.
The consistency condition of (\ref{Lax})
\be
 L_{p}(\hat{\bv},\hat{\til{\bv}})L_{q}(\bv,\hat{\bv})-L_{q}(\til{\bv},\hat{\til{\bv}})L_{p}(\bv,\til{\bv})=\bze
\ee
gives the system \eqref{discreteMGD}.

The multidimensional consistency of (\ref{discreteMGD}), like the Lax pair, is basically inherited from (\ref{notHirotaMiwa}).
For this property, the vector $\bv=\bv(l,m)$ on $\mathbb{Z}^2$ is considered instead as a function on the $d$-dimensional regular lattice $\bv=\bv(l_1,l_2,\ldots,l_d)$ satisfying system (\ref{discreteMGD}) in all pairs of directions, where parameters $p,q$ appearing in (\ref{discreteMGD}) associated with the original $l$ and $m$ directions are to be replaced with pairs of parameters taken from the larger set $p_1,\ldots,p_d$ associated with the corresponding $l_1,\ldots,l_d$ directions.
The consistency of this system can be verified directly by calculation.

\subsection{Point symmetries}\label{Section_Symmetry}
The system (\ref{discreteMGD}) is invariant under the scaling symmetry
\be
(v_1,\ldots,v_N)\rightarrow(\mu_1 v_1,\ldots, \mu_{N} v_{N}), \qquad \mu_1,\ldots,\mu_N \in\mathbb{C}\setminus\{0\},\label{sym1}
\ee
as well as the less obvious transformation
\be
(v_1,v_2,\ldots,v_{N-1},v_{N})\rightarrow (v_2/v_1,v_3/v_1,\ldots,v_N/v_{1},1/v_1).\label{sym2}
\ee
The latter transformation here generates a group of $N$ similar symmetries of (\ref{discreteMGD}).
Such autonomous point symmetries allow an equation to take on alternative guises via a non-autonomous point transformation that preserves the autonomous nature of the equation, and also preserves its multidimensional consistency property (see for instance \cite{BobSur2002,Atk2009}).
Explicitly, if $S_1$ and $S_2$ are (commuting) transformations of the vector $\bv=(v_1,\ldots,v_N)$ taken from the group generated by (\ref{sym1}) and (\ref{sym2}), and $\bv$ is governed by (\ref{discreteMGD}), then the equation governing $\bv'=[S_1^l\cdot S_2^m](\bv)$ is both autonomous and multidimensionally consistent.
The autonomous nature of the equation is preserved due to the fact the $S_i$ are symmetries, whilst the natural extension to multidimensions of the transformation: $[S_1^l\cdot S_2^m] \rightarrow [S_1^{l_1}\cdot S_2^{l_2} \cdot S_3^{l_3}\cdot \ldots ]$, demonstrates also that the consistency is preserved.

This is an important part of the transformation theory of multi-component systems such as (\ref{discreteMGD}) because invariants of such transformations are not presently known beyond the multi-affine class \cite{AdlBobSur2009a}.
The less obvious symmetry (\ref{sym2}) is included here because to distinguish between genuinely different quad equations in the multi-component class, it is necessary to be aware of all such transformations.

\section{Lagrangian structure}\label{Section_Lagrangian}
For discrete systems where each lattice direction is on an equal footing it appears that the Lagrangian is a more natural object than the Hamiltonian, given that it does not single out any one direction. Therefore an important feature of the multicomponent quad equation \eqref{discreteMGD} is its Lagrangian structure, which as we will see is inherited from that of Hirota's discrete KP equation \eqref{notHirotaMiwa}. The more remarkable feature is its natural interplay with the multidimensional consistency of the model, which will be explained in the following section (Section \ref{Section_closure}). 

A Lagrangian for the system \eqref{discreteMGD} is
\bea\label{DiscreteMGDLagrangian}
\cL_{pq} & = & \sum_{i=0}^{N}\biggl\{\Li_{2}\biggl(\frac{q\til{v}_{i}\hat{v}_{i+1}}{p\hat{v}_{i}\til{v}_{i+1}}\biggr) 
                                     + \frac{1}{4}\biggl(\ln\biggl(-\frac{q\til{v}_{i}\hat{v}_{i+1}}{p\hat{v}_{i}\til{v}_{i+1}}\biggr)\biggr)^2
                                     + \frac{1}{2}\bigl(\ln{\hat{v}_{i}}\ln{\til{v}_{i+1}}-\ln{\til{v}_{i}}\ln{\hat{v}_{i+1}}\bigr)\nn\\
            && \;\;\;\;\;\;\;\;\;    + \ln\biggl(\frac{v_{i+1}}{v_{i}}\biggr)\ln\biggl(\frac{\til{v}_{i}}{\hat{v}_{i}}\biggr)\biggr\},
\eea
which through the discrete Euler-Lagrange equations gives 2 copies of the equations. The function $\Li_{2}$ is the dilogarithm function
\be\label{Li2}
 \Li_2(z) = -\int^z_0{\frac{\ln(1-z)}{z}dz},
\ee
which appears in many areas of physics, such as quantum electrodynamics (e.g. vacuum polarization) and electrical network problems; it also comes up in connection with algebraic K-theory, representation theory of infinite dimensional algebras, and combinatorics \cite{Kir1995}. 

By discrete Euler-Lagrange equations we mean in this case
\be
 \frac{\dl\cL_{pq}}{\dl v_{i}}\equiv \frac{\ptl\cL_{pq}}{\ptl v_{i}}+T_{p}^{-1}\frac{\ptl\cL_{pq}}{\ptl \til{v}_{i}}+T_{q}^{-1}\frac{\ptl\cL_{pq}}{\ptl \hat{v}_{i}} = 0, \;\;\;\; i\in\{1,\dots,N\},
\ee
where $T_{p}$ is a shift operator in the lattice direction associated with parameter $p$, and similarly $T_{q}$ is a shift operator in the lattice direction associated with parameter $q$. Then $T_{p}^{-1},T_{q}^{-1}$ are simply backwards shifts in the lattice.

The discrete Euler-Lagrange equations arise from extremizing the action
\bea\label{DiscreteMGDaction}
 S & = & \sum_{l}\sum_{m}\sum_{i=0}^{N}\biggl\{\Li_{2}\biggl(\frac{q\til{v}_{i}\hat{v}_{i+1}}{p\hat{v}_{i}\til{v}_{i+1}}\biggr) 
                                     + \frac{1}{4}\biggl(\ln\biggl(-\frac{q\til{v}_{i}\hat{v}_{i+1}}{p\hat{v}_{i}\til{v}_{i+1}}\biggr)\biggr)^2
                                     + \frac{1}{2}\bigl(\ln{\hat{v}_{i}}\ln{\til{v}_{i+1}}-\ln{\til{v}_{i}}\ln{\hat{v}_{i+1}}\bigr)\nn\\
            && \;\;\;\;\;\;\;\;\;    + \ln\biggl(\frac{v_{i+1}}{v_{i}}\biggr)\ln\biggl(\frac{\til{v}_{i}}{\hat{v}_{i}}\biggr)\biggr\},
\eea
where the first two sums over $l$ and $m$ are understood to be taken over the whole 2-dimensional lattice. In fact, rather than the Lagrangian by itself, it is more instructive to look at the action, since it is on this level that we can easily see we are dealing with a reduction of the bilinear discrete KP equation
\be\label{HM}
 A_{pq}\bar{\tau}\hat{\til{\tau}}+A_{qr}\til{\tau}\hat{\bar{\tau}}+A_{rp}\hat{\tau}\til{\bar{\tau}}=0;
\ee
here the dependent variable is $\tau$, and the $A_{pq}$ are antisymmetric constants (so that $A_{qp}=-A_{pq}$) associated with the $p,q$ lattice directions. Making the identification $A_{pr}/A_{pq}=p$, $A_{qr}/A_{pq}=q$ and $\tau=u$, so that the equation is no longer covariant due to the singled-out lattice direction, brings us to equation \eqref{notHirotaMiwa}.
A Lagrangian for \eqref{HM} was presented in \cite{LobNijQui2009}, which, adapted for the non-covariant equation \eqref{notHirotaMiwa}, is equivalent to 
\bea
 S_{KP} & = & \sum_{l}\sum_{m}\sum_{n}\biggl\{\Li_{2}\biggl(\frac{q\til{\tau}\hat{\bar{\tau}}}{p\hat{\tau}\til{\bar{\tau}}}\biggr) 
                                     + \frac{1}{4}\biggl(\ln\biggl(-\frac{q\til{\tau}\hat{\bar{\tau}}}{p\hat{\tau}\til{\bar{\tau}}}\biggr)\biggr)^2
                                     + \frac{1}{2}\bigl(\ln{\hat{\tau}}\ln{\til{\bar{\tau}}}-\ln{\til{\tau}}\ln{\hat{\bar{\tau}}}\bigr)\nn\\
            && \;\;\;\;\;\;\;\;\;    + \ln\biggl(\frac{\bar{\tau}}{\tau}\biggr)\ln\biggl(\frac{\til{\tau}}{\hat{\tau}}\biggr)\biggr\}.
\eea
It is clear that identifying $\tau=v_{i}$, $\bar{\tau}=v_{i+1}$, and restricting the sum over $n$, we get the action \eqref{DiscreteMGDaction} of the reduction.

\section{Closure property}\label{Section_closure}

The Lagrangians for many integrable systems can be interpreted as closed forms; they obey a closure relation on solutions when consistently embedded in a higher-dimensional space. In a sense one can consider this to be multidimensional consistency on the level of the Lagrangian. It allows for a variational principle which includes the geometry of the space of independent variables, as well as that of the dependent variables; seeking extrema of the action under variations with respect to all of the variables, dependent and independent, leads to constraints which specify not only the equations of motion, but also to an extent the Lagrangians themselves. 

In the case of 2-dimensional systems, we have Lagrangian 2-forms, for instance Lagrangian functions which are evaluated on 2-dimensional plaquettes. For multidimensionally consistent systems we can suppose the action to be defined on a surface in an arbitrary number of dimensions, not just two, and this action will be a sum of the Lagrangian function evaluated on every plaquette in that surface. Demanding zero variation with respect to the surface implies a closure relation on the Lagrangian, while variation with respect to the dependent variable gives equations of motion.

Examples can be found in both discrete and continuous cases of 1-dimensional systems \cite{YooLobNij2011,YooNij2011}, a large class of 2-dimensional systems \cite{LobNij2009,LobNij2010,XenNijLob2010}, and also the 3-dimensional discrete bilinear KP equation \cite{LobNijQui2009}. In \cite{LobNij2010} it was demonstrated that the Lagrangian for an arbitrary member of the lattice GD hierarchy obeys the closure relation
\be\label{closure}
 \Dl_{p}\cL_{qr}+\Dl_{q}\cL_{rp}+\Dl_{r}\cL_{pq} = 0,
\ee
on solutions to the lattice KP equation (the lattice KP equation is a consequence of the equations which constitute the lattice GD hierarchy itself). Here the operator $\Dl_{p}$ is a difference operator in the $p$-direction of the lattice; if we again employ $T_{p}$ to denote the shift operator in the $p$-direction, and write the identity operator as $id$, then $\Dl_{p}=T_{p}-id$.

To demonstrate the Lagrangian \eqref{DiscreteMGDLagrangian} satisfies the closure relation \eqref{closure}, we need to invoke a 5-term identity for the dilogarithm function
\bea\label{5pt}
 \Li_2(s) + \Li_2(t) - \Li_2(st)& = & \Li_2\biggl(\frac{s-st}{1-st}\biggr)+\Li_2\biggl(\frac{t-st}{1-st}\biggr)\nn\\
                                   && +\ln\biggl(\frac{1-s}{1-st}\biggr)\ln\biggl(\frac{1-t}{1-st}\biggr),
\eea
which holds for $s,t,st\neq 1$, up to imaginary constant terms, and also a 2-term identity which flips the argument
\be\label{flip}
  \Li_2(s)+\Li_2\biggl(\frac{1}{s}\biggr) = -\frac{1}{2}\bigl(\ln(-s)\bigr)^2-\frac{\pi^2}{6},
\ee
for $s\neq 0$. Further identities and other information about the dilogarithm function can be found in \cite{Lew1981}.

Observe that

\setlength{\unitlength}{1cm}
\begin{picture}(10,9)(0,0)
 \put(0,9){\parbox[t]{6cm}{
\bea
\Gamma & \equiv & \bar{\cL}_{pq}+\til{\cL}_{qr}+\hat{\cL}_{rp}-\cL_{pq}-\cL_{qr}-\cL_{rp}\nn\\
            & = & \sum_{i=0}^{N}\biggl\{\;\Li_{2}\biggl(\frac{q\til{\bar{v}}_{i}\hat{\bar{v}}_{i+1}}{p\hat{\bar{v}}_{i}\til{\bar{v}}_{i+1}}\biggr)\;\;
                                       +\Li_{2}\biggl(\frac{r\hat{\til{v}}_{i}\til{\bar{v}}_{i+1}}{q\til{\bar{v}}_{i}\hat{\til{v}}_{i+1}}\biggr)
                                       +\Li_{2}\biggl(\frac{p\hat{\bar{v}}_{i}\hat{\til{v}}_{i+1}}{r\hat{\til{v}}_{i}\hat{\bar{v}}_{i+1}}\biggr)\nn\\
               && \;\;\;\;\;\;\;       -\Li_{2}\biggl(\frac{q\til{v}_{i}\hat{v}_{i+1}}{p\hat{v}_{i}\til{v}_{i+1}}\biggr)\;\;
                                       -\Li_{2}\biggl(\frac{r\hat{v}_{i}\bar{v}_{i+1}}{q\bar{v}_{i}\hat{v}_{i+1}}\biggr)
                                       -\Li_{2}\biggl(\frac{p\bar{v}_{i}\til{v}_{i+1}}{r\til{v}_{i}\bar{v}_{i+1}}\biggr)\nn\\
               && \;\;\;\;\;\;\;        + \frac{1}{4}\biggl(\ln\biggl(-\frac{q\til{\bar{v}}_{i}\hat{\bar{v}}_{i+1}}{p\hat{\bar{v}}_{i}\til{\bar{v}}_{i+1}}\biggr)\biggr)^2
                                        + \frac{1}{4}\biggl(\ln\biggl(-\frac{r\hat{\til{v}}_{i}\til{\bar{v}}_{i+1}}{q\til{\bar{v}}_{i}\hat{\til{v}}_{i+1}}\biggr)\biggr)^2
                                        + \frac{1}{4}\biggl(\ln\biggl(-\frac{p\hat{\bar{v}}_{i}\hat{\til{v}}_{i+1}}{r\hat{\til{v}}_{i}\hat{\bar{v}}_{i+1}}\biggr)\biggr)^2\nn\\
               && \;\;\;\;\;\;\;        - \frac{1}{4}\biggl(\ln\biggl(-\frac{q\til{v}_{i}\hat{v}_{i+1}}{p\hat{v}_{i}\til{v}_{i+1}}\biggr)\biggr)^2
                                        - \frac{1}{4}\biggl(\ln\biggl(-\frac{r\hat{v}_{i}\bar{v}_{i+1}}{q\bar{v}_{i}\hat{v}_{i+1}}\biggr)\biggr)^2
                                        - \frac{1}{4}\biggl(\ln\biggl(-\frac{p\bar{v}_{i}\til{v}_{i+1}}{r\til{v}_{i}\bar{v}_{i+1}}\biggr)\biggr)^2\nn\\
               && \;\;\;\;\;\;\;        + \frac{1}{2}\bigl(\ln{\hat{\bar{v}}_{i}}\ln{\til{\bar{v}}_{i+1}}-\ln{\til{\bar{v}}_{i}}\ln{\hat{\bar{v}}_{i+1}}\bigr)
                                        + \frac{1}{2}\bigl(\ln{\til{\bar{v}}_{i}}\ln{\hat{\til{v}}_{i+1}}-\ln{\hat{\til{v}}_{i}}\ln{\til{\bar{v}}_{i+1}}\bigr)\nn\\
               && \;\;\;\;\;\;\;        + \frac{1}{2}\bigl(\ln{\hat{\til{v}}_{i}}\ln{\hat{\bar{v}}_{i+1}}-\ln{\hat{\bar{v}}_{i}}\ln{\hat{\til{v}}_{i+1}}\bigr)
                                        - \frac{1}{2}\bigl(\ln{\hat{v}_{i}}\ln{\til{v}_{i+1}}-\ln{\til{v}_{i}}\ln{\hat{v}_{i+1}}\bigr)\nn\\
               && \;\;\;\;\;\;\;        - \frac{1}{2}\bigl(\ln{\bar{v}_{i}}\ln{\hat{v}_{i+1}}-\ln{\hat{v}_{i}}\ln{\bar{v}_{i+1}}\bigr)
                                        - \frac{1}{2}\bigl(\ln{\til{v}_{i}}\ln{\bar{v}_{i+1}}-\ln{\bar{v}_{i}}\ln{\til{v}_{i+1}}\bigr)\nn\\
               && \;\;\;\;\;\;\;        + \ln\biggl(\frac{\bar{v}_{i+1}}{\bar{v}_{i}}\biggr)\ln\biggl(\frac{\til{\bar{v}}_{i}}{\hat{\bar{v}}_{i}}\biggr)
                                        + \ln\biggl(\frac{\til{v}_{i+1}}{\til{v}_{i}}\biggr)\ln\biggl(\frac{\hat{\til{v}}_{i}}{\til{\bar{v}}_{i}}\biggr)
                                        + \ln\biggl(\frac{\hat{v}_{i+1}}{\hat{v}_{i}}\biggr)\ln\biggl(\frac{\hat{\bar{v}}_{i}}{\hat{\til{v}}_{i}}\biggr)\nn\\
               && \;\;\;\;\;\;\;        - \ln\biggl(\frac{v_{i+1}}{v_{i}}\biggr)\ln\biggl(\frac{\til{v}_{i}}{\hat{v}_{i}}\biggr)
                                        - \ln\biggl(\frac{v_{i+1}}{v_{i}}\biggr)\ln\biggl(\frac{\hat{v}_{i}}{\bar{v}_{i}}\biggr)
                                        - \ln\biggl(\frac{v_{i+1}}{v_{i}}\biggr)\ln\biggl(\frac{\bar{v}_{i}}{\til{v}_{i}}\biggr)\biggr\}.
\eea
 }}
 \put(1.9,7.2){\dashbox{0.1}(2.1,0.9)}\put(4.1,7.2){\framebox(5,0.9)}
 \put(1.7,6.2){\dashbox{0.1}(2.4,0.8)}\put(4.2,6.2){\framebox(5,0.8)}
\end{picture}

Using the 2-term dilogarithm identity \eqref{flip} on the terms in the dashed-line boxes, and the 5-point identity \eqref{5pt} on the terms in the solid-line boxes, this becomes:

\bea
\Gamma & = & \sum_{i=0}^{N}\biggl\{\Li_{2}\biggl(-\frac{p\hat{\bar{v}}_{i}}{r\hat{\til{v}}_{i}}\cdot
                                                  \frac{q\til{\bar{v}}_{i}\hat{\til{v}}_{i+1}-r\hat{\til{v}}_{i}\til{\bar{v}}_{i+1}}{p\hat{\bar{v}}_{i}\til{\bar{v}}_{i+1}-q\til{\bar{v}}_{i}\hat{\bar{v}}_{i+1}}\biggr)
                                  +\Li_{2}\biggl(-\frac{\til{\bar{v}}_{i+1}}{\hat{\til{v}}_{i+1}}\cdot
                                                  \frac{r\hat{\til{v}}_{i}\hat{\bar{v}}_{i+1}-p\hat{\bar{v}}_{i}\hat{\til{v}}_{i+1}}{p\hat{\bar{v}}_{i}\til{\bar{v}}_{i+1}-q\til{\bar{v}}_{i}\hat{\bar{v}}_{i+1}}\biggr)\nn\\
          && \;\;\;\;\;\;\;       -\Li_{2}\biggl(-\frac{p\til{v}_{i+1}}{r\bar{v}_{i+1}}\cdot
                                                  \frac{q\bar{v}_{i}\hat{v}_{i+1}-r\hat{v}_{i}\bar{v}_{i+1}}{p\hat{v}_{i}\til{v}_{i+1}-q\til{v}_{i}\hat{v}_{i+1}}\biggr)
                                  -\Li_{2}\biggl(-\frac{\hat{v}_{i}}{\bar{v}_{i}}\cdot
                                                  \frac{r\til{v}_{i}\bar{v}_{i+1}-p\bar{v}_{i}\til{v}_{i+1}}{p\hat{v}_{i}\til{v}_{i+1}-q\til{v}_{i}\hat{v}_{i+1}}\biggr)\nn\\
          && \;\;\;\;\;\;\;        + \ln\biggl(-\frac{\hat{\bar{v}}_{i+1}}{\hat{\til{v}}_{i+1}}\cdot
                                                \frac{q\til{\bar{v}}_{i}\hat{\til{v}}_{i+1}-r\hat{\til{v}}_{i}\til{\bar{v}}_{i+1}}{p\hat{\bar{v}}_{i}\til{\bar{v}}_{i+1}-q\til{\bar{v}}_{i}\hat{\bar{v}}_{i+1}}\biggr)
                                     \ln\biggl(-\frac{q\til{\bar{v}}_{i}}{r\hat{\til{v}}_{i}}\cdot
                                                \frac{r\hat{\til{v}}_{i}\hat{\bar{v}}_{i+1}-p\hat{\bar{v}}_{i}\hat{\til{v}}_{i+1}}{p\hat{\bar{v}}_{i}\til{\bar{v}}_{i+1}-q\til{\bar{v}}_{i}\hat{\bar{v}}_{i+1}}\biggr)\nn\\
          && \;\;\;\;\;\;\;        - \ln\biggl(-\frac{\til{v}_{i}}{\bar{v}_{i}}\cdot
                                                \frac{q\bar{v}_{i}\hat{v}_{i+1}-r\hat{v}_{i}\bar{v}_{i+1}}{p\hat{v}_{i}\til{v}_{i+1}-q\til{v}_{i}\hat{v}_{i+1}}\biggr)
                                     \ln\biggl(-\frac{q\hat{v}_{i+1}}{r\bar{v}_{i+1}}\cdot
                                                  \frac{r\til{v}_{i}\bar{v}_{i+1}-p\bar{v}_{i}\til{v}_{i+1}}{p\hat{v}_{i}\til{v}_{i+1}-q\til{v}_{i}\hat{v}_{i+1}}\biggr)\nn\\
          && \;\;\;\;\;\;\;        - \frac{1}{4}\biggl(\ln\biggl(-\frac{q\til{\bar{v}}_{i}\hat{\bar{v}}_{i+1}}{p\hat{\bar{v}}_{i}\til{\bar{v}}_{i+1}}\biggr)\biggr)^2
                                   + \frac{1}{4}\biggl(\ln\biggl(-\frac{r\hat{\til{v}}_{i}\til{\bar{v}}_{i+1}}{q\til{\bar{v}}_{i}\hat{\til{v}}_{i+1}}\biggr)\biggr)^2
                                   + \frac{1}{4}\biggl(\ln\biggl(-\frac{p\hat{\bar{v}}_{i}\hat{\til{v}}_{i+1}}{r\hat{\til{v}}_{i}\hat{\bar{v}}_{i+1}}\biggr)\biggr)^2\nn\\
          && \;\;\;\;\;\;\;        + \frac{1}{4}\biggl(\ln\biggl(-\frac{q\til{v}_{i}\hat{v}_{i+1}}{p\hat{v}_{i}\til{v}_{i+1}}\biggr)\biggr)^2
                                   - \frac{1}{4}\biggl(\ln\biggl(-\frac{r\hat{v}_{i}\bar{v}_{i+1}}{q\bar{v}_{i}\hat{v}_{i+1}}\biggr)\biggr)^2
                                   - \frac{1}{4}\biggl(\ln\biggl(-\frac{p\bar{v}_{i}\til{v}_{i+1}}{r\til{v}_{i}\bar{v}_{i+1}}\biggr)\biggr)^2\nn\\
          && \;\;\;\;\;\;\;        + \frac{1}{2}\bigl(\ln{\hat{\bar{v}}_{i}}\ln{\til{\bar{v}}_{i+1}}-\ln{\til{\bar{v}}_{i}}\ln{\hat{\bar{v}}_{i+1}}\bigr)
                                   + \frac{1}{2}\bigl(\ln{\til{\bar{v}}_{i}}\ln{\hat{\til{v}}_{i+1}}-\ln{\hat{\til{v}}_{i}}\ln{\til{\bar{v}}_{i+1}}\bigr)\nn\\
          && \;\;\;\;\;\;\;        + \frac{1}{2}\bigl(\ln{\hat{\til{v}}_{i}}\ln{\hat{\bar{v}}_{i+1}}-\ln{\hat{\bar{v}}_{i}}\ln{\hat{\til{v}}_{i+1}}\bigr)
                                   - \frac{1}{2}\bigl(\ln{\hat{v}_{i}}\ln{\til{v}_{i+1}}-\ln{\til{v}_{i}}\ln{\hat{v}_{i+1}}\bigr)\nn\\
          && \;\;\;\;\;\;\;        - \frac{1}{2}\bigl(\ln{\bar{v}_{i}}\ln{\hat{v}_{i+1}}-\ln{\hat{v}_{i}}\ln{\bar{v}_{i+1}}\bigr)
                                   - \frac{1}{2}\bigl(\ln{\til{v}_{i}}\ln{\bar{v}_{i+1}}-\ln{\bar{v}_{i}}\ln{\til{v}_{i+1}}\bigr)\nn\\
          && \;\;\;\;\;\;\;        + \ln\biggl(\frac{\bar{v}_{i+1}}{\bar{v}_{i}}\biggr)\ln\biggl(\frac{\til{\bar{v}}_{i}}{\hat{\bar{v}}_{i}}\biggr)
                                   + \ln\biggl(\frac{\til{v}_{i+1}}{\til{v}_{i}}\biggr)\ln\biggl(\frac{\hat{\til{v}}_{i}}{\til{\bar{v}}_{i}}\biggr)
                                   + \ln\biggl(\frac{\hat{v}_{i+1}}{\hat{v}_{i}}\biggr)\ln\biggl(\frac{\hat{\bar{v}}_{i}}{\hat{\til{v}}_{i}}\biggr)\biggr\}.
\eea

From the equations, we have that 

\bse
\bea
 \hat{\til{v}}_{i}  & = & \frac{p\hat{v}_{i}\til{v}_{i+1}-q\til{v}_{i}\hat{v}_{i+1}}{p\til{v}_{1}-q\hat{v}_{1}}\cdot\frac{v_{1}}{v_{i+1}},\\
 \hat{\bar{v}}_{i}  & = & \frac{q\bar{v}_{i}\hat{v}_{i+1}-r\hat{v}_{i}\bar{v}_{i+1}}{q\hat{v}_{1}-r\bar{v}_{1}}\cdot\frac{v_{1}}{v_{i+1}},\\
 \til{\bar{v}}_{i}  & = & \frac{r\til{v}_{i}\bar{v}_{i+1}-p\bar{v}_{i}\til{v}_{i+1}}{r\bar{v}_{1}-p\til{v}_{1}}\cdot\frac{v_{1}}{v_{i+1}},
\eea
\ese
for $i\in\{0,\dots,N\}$, and of course $v_{0}=v_{N+1}=1$ (so for $i=1$ these are trivial identities). Identifying also $v_{N+2}=v_{1}$, the equations hold trivially for $i=N+1$. Substituting these in gives

\bea
\Gamma & = & \sum_{i=0}^{N}\biggl\{\Li_{2}\biggl(-\frac{p\til{v}_{i+1}}{r\bar{v}_{i+1}}\cdot
                                                  \frac{q\bar{v}_{i}\hat{v}_{i+1}-r\hat{v}_{i}\bar{v}_{i+1}}{p\hat{v}_{i}\til{v}_{i+1}-q\til{v}_{i}\hat{v}_{i+1}}\biggr)
                                  +\Li_{2}\biggl(-\frac{\hat{v}_{i+1}}{\bar{v}_{i+1}}\cdot
                                                  \frac{r\til{v}_{i+1}\bar{v}_{i+2}-p\bar{v}_{i+1}\til{v}_{i+2}}{p\hat{v}_{i+1}\til{v}_{i+2}-q\til{v}_{i+1}\hat{v}_{i+2}}\biggr)\nn\\
          && \;\;\;\;\;\;\;       -\Li_{2}\biggl(-\frac{p\til{v}_{i+1}}{r\bar{v}_{i+1}}\cdot
                                                  \frac{q\bar{v}_{i}\hat{v}_{i+1}-r\hat{v}_{i}\bar{v}_{i+1}}{p\hat{v}_{i}\til{v}_{i+1}-q\til{v}_{i}\hat{v}_{i+1}}\biggr)
                                  -\Li_{2}\biggl(-\frac{\hat{v}_{i}}{\bar{v}_{i}}\cdot
                                                  \frac{r\til{v}_{i}\bar{v}_{i+1}-p\bar{v}_{i}\til{v}_{i+1}}{p\hat{v}_{i}\til{v}_{i+1}-q\til{v}_{i}\hat{v}_{i+1}}\biggr)\nn\\
          && \;\;\;\;\;\;\;        + \ln\biggl(-\frac{\til{v}_{i+1}}{\bar{v}_{i+1}}\cdot
                                                \frac{q\bar{v}_{i}\hat{v}_{i+1}-r\hat{v}_{i}\bar{v}_{i+1}}{p\hat{v}_{i}\til{v}_{i+1}-q\til{v}_{i}\hat{v}_{i+1}}\biggr)
                                     \ln\biggl(-\frac{q\hat{v}_{i+1}}{r\bar{v}_{i+1}}\cdot
                                                  \frac{r\til{v}_{i+1}\bar{v}_{i+2}-p\bar{v}_{i+1}\til{v}_{i+2}}{p\hat{v}_{i+1}\til{v}_{i+2}-q\til{v}_{i+1}\hat{v}_{i+2}}\biggr)\nn\\
          && \;\;\;\;\;\;\;        - \ln\biggl(-\frac{\til{v}_{i}}{\bar{v}_{i}}\cdot
                                                \frac{q\bar{v}_{i}\hat{v}_{i+1}-r\hat{v}_{i}\bar{v}_{i+1}}{p\hat{v}_{i}\til{v}_{i+1}-q\til{v}_{i}\hat{v}_{i+1}}\biggr)
                                     \ln\biggl(-\frac{q\hat{v}_{i+1}}{r\bar{v}_{i+1}}\cdot
                                                  \frac{r\til{v}_{i}\bar{v}_{i+1}-p\bar{v}_{i}\til{v}_{i+1}}{p\hat{v}_{i}\til{v}_{i+1}-q\til{v}_{i}\hat{v}_{i+1}}\biggr)\nn\\
          && \;\;\;\;\;\;\;        - \frac{1}{4}\biggl(\ln\biggl(-\frac{q\til{\bar{v}}_{i}\hat{\bar{v}}_{i+1}}{p\hat{\bar{v}}_{i}\til{\bar{v}}_{i+1}}\biggr)\biggr)^2
                                   + \frac{1}{4}\biggl(\ln\biggl(-\frac{r\hat{\til{v}}_{i}\til{\bar{v}}_{i+1}}{q\til{\bar{v}}_{i}\hat{\til{v}}_{i+1}}\biggr)\biggr)^2
                                   + \frac{1}{4}\biggl(\ln\biggl(-\frac{p\hat{\bar{v}}_{i}\hat{\til{v}}_{i+1}}{r\hat{\til{v}}_{i}\hat{\bar{v}}_{i+1}}\biggr)\biggr)^2\nn\\
          && \;\;\;\;\;\;\;        + \frac{1}{4}\biggl(\ln\biggl(-\frac{q\til{v}_{i}\hat{v}_{i+1}}{p\hat{v}_{i}\til{v}_{i+1}}\biggr)\biggr)^2
                                   - \frac{1}{4}\biggl(\ln\biggl(-\frac{r\hat{v}_{i}\bar{v}_{i+1}}{q\bar{v}_{i}\hat{v}_{i+1}}\biggr)\biggr)^2
                                   - \frac{1}{4}\biggl(\ln\biggl(-\frac{p\bar{v}_{i}\til{v}_{i+1}}{r\til{v}_{i}\bar{v}_{i+1}}\biggr)\biggr)^2\nn\\
          && \;\;\;\;\;\;\;        + \frac{1}{2}\bigl(\ln{\hat{\bar{v}}_{i}}\ln{\til{\bar{v}}_{i+1}}-\ln{\til{\bar{v}}_{i}}\ln{\hat{\bar{v}}_{i+1}}\bigr)
                                   + \frac{1}{2}\bigl(\ln{\til{\bar{v}}_{i}}\ln{\hat{\til{v}}_{i+1}}-\ln{\hat{\til{v}}_{i}}\ln{\til{\bar{v}}_{i+1}}\bigr)\nn\\
          && \;\;\;\;\;\;\;        + \frac{1}{2}\bigl(\ln{\hat{\til{v}}_{i}}\ln{\hat{\bar{v}}_{i+1}}-\ln{\hat{\bar{v}}_{i}}\ln{\hat{\til{v}}_{i+1}}\bigr)
                                   - \frac{1}{2}\bigl(\ln{\hat{v}_{i}}\ln{\til{v}_{i+1}}-\ln{\til{v}_{i}}\ln{\hat{v}_{i+1}}\bigr)\nn\\
          && \;\;\;\;\;\;\;        - \frac{1}{2}\bigl(\ln{\bar{v}_{i}}\ln{\hat{v}_{i+1}}-\ln{\hat{v}_{i}}\ln{\bar{v}_{i+1}}\bigr)
                                   - \frac{1}{2}\bigl(\ln{\til{v}_{i}}\ln{\bar{v}_{i+1}}-\ln{\bar{v}_{i}}\ln{\til{v}_{i+1}}\bigr)\nn\\
          && \;\;\;\;\;\;\;        + \ln\biggl(\frac{\bar{v}_{i+1}}{\bar{v}_{i}}\biggr)\ln\biggl(\frac{\til{\bar{v}}_{i}}{\hat{\bar{v}}_{i}}\biggr)
                                   + \ln\biggl(\frac{\til{v}_{i+1}}{\til{v}_{i}}\biggr)\ln\biggl(\frac{\hat{\til{v}}_{i}}{\til{\bar{v}}_{i}}\biggr)
                                   + \ln\biggl(\frac{\hat{v}_{i+1}}{\hat{v}_{i}}\biggr)\ln\biggl(\frac{\hat{\bar{v}}_{i}}{\hat{\til{v}}_{i}}\biggr)\biggr\}.
\eea

Clearly some of these terms will cancel, reducing to

\bea
\Gamma & = & \sum_{i=0}^{N}\frac{1}{2}\biggl\{ 2\biggl[\Li_{2}\biggl(-\frac{\hat{v}_{i+1}}{\bar{v}_{i+1}}\cdot
                                                  \frac{r\til{v}_{i+1}\bar{v}_{i+2}-p\bar{v}_{i+1}\til{v}_{i+2}}{p\hat{v}_{i+1}\til{v}_{i+2}-q\til{v}_{i+1}\hat{v}_{i+2}}\biggr)
                                   -\Li_{2}\biggl(-\frac{\hat{v}_{i}}{\bar{v}_{i}}\cdot
                                                  \frac{r\til{v}_{i}\bar{v}_{i+1}-p\bar{v}_{i}\til{v}_{i+1}}{p\hat{v}_{i}\til{v}_{i+1}-q\til{v}_{i}\hat{v}_{i+1}}\biggr) \biggr]\nn\\
          && \;\;\;\;\;\;\;        + 2\ln\biggl(-\frac{r\bar{v}_{1}-p\til{v}_{1}}{p\til{v}_{1}-q\hat{v}_{1}}\biggr)
                                      \biggl[\ln\biggl(\frac{\hat{\bar{v}}_{i+1}\til{v}_{i+1}}{\hat{\til{v}}_{i+1}\bar{v}_{i+1}}\biggr)
                                            -\ln\biggl(\frac{\hat{\bar{v}}_{i}\til{v}_{i}}{\hat{\til{v}}_{i}\bar{v}_{i}}\biggr)\biggr]\nn\\
          && \;\;\;\;\;\;\;        - \ln{p}\biggl[\ln\biggl(\frac{\til{\bar{v}}_{i+1}\hat{v}_{i+1}}{\hat{\til{v}}_{i+1}\bar{v}_{i+1}}\biggr)
                                           -\ln\biggl(\frac{\til{\bar{v}}_{i}\hat{v}_{i}}{\hat{\til{v}}_{i}\bar{v}_{i}}\biggr)\biggr]
                                   - \ln{q}\biggl[\ln\biggl(\frac{\hat{\bar{v}}_{i+1}\til{v}_{i+1}}{\hat{\til{v}}_{i+1}\bar{v}_{i+1}}\biggr)
                                           -\ln\biggl(\frac{\hat{\bar{v}}_{i}\til{v}_{i}}{\hat{\til{v}}_{i}\bar{v}_{i}}\biggr)\biggr]\nn\\
          && \;\;\;\;\;\;\;        + \ln{r}\bigl[\ln\bigl(\hat{\bar{v}}_{i+1}\til{\bar{v}}_{i+1}\til{v}_{i+1}\hat{v}_{i+1}\bigr)
                                           -\ln\bigl(\hat{\bar{v}}_{i}\til{\bar{v}}_{i}\til{v}_{i}\hat{v}_{i}\bigr)\bigr]
                                   - 2\ln{r}\bigl[\ln\bigl(\hat{\til{v}}_{i+1}\bar{v}_{i+1}\bigr)
                                            -\ln\bigl(\hat{\til{v}}_{i}\bar{v}_{i}\bigr)\bigr]\nn\\
         && \;\;\;\;\;\;\; +\bigl[(\ln{\hat{\til{v}}_{i+1}})^2-(\ln{\hat{\til{v}}_{i}})^2\bigr]-\bigl[\ln{\til{\bar{v}}_{i+1}}\ln{\hat{\til{v}}_{i+1}}-\ln{\hat{\til{v}}_{i}}\ln{\til{\bar{v}}_{i}}\bigr]
                           +\bigl[\ln{\hat{\bar{v}}_{i+1}}\ln{\til{\bar{v}}_{i+1}}-\ln{\til{\bar{v}}_{i}}\ln{\hat{\bar{v}}_{i}}\bigr]\nn\\
         && \;\;\;\;\;\;\; +2\bigl[\ln{\bar{v}_{i+1}}\ln{\hat{\til{v}}_{i+1}}-\ln{\hat{\til{v}}_{i}}\ln{\bar{v}_{i}}\bigr]
                           -2\bigl[\ln{\bar{v}_{i+1}}\ln{\hat{\bar{v}}_{i+1}}-\ln{\bar{v}_{i}}\ln{\hat{\bar{v}}_{i}}\bigr]\nn\\
         && \;\;\;\;\;\;\; +2\bigl[\ln{\hat{v}_{i+1}}\ln{\hat{\bar{v}}_{i+1}}-\ln{\hat{v}_{i}}\ln{\hat{\bar{v}}_{i}}\bigr]
                           -2\bigl[\ln{\hat{v}_{i+1}}\ln{\hat{\til{v}}_{i+1}}-\ln{\hat{v}_{i}}\ln{\hat{\til{v}}_{i}}\bigr]\nn\\
         && \;\;\;\;\;\;\; -\bigl[\ln{\til{v}_{i+1}}\ln{\bar{v}_{i+1}}-\ln{\bar{v}_{i}}\ln{\til{v}_{i}}\bigr]+\bigl[\ln{\hat{v}_{i+1}}\ln{\til{v}_{i+1}}-\ln{\til{v}_{i}}\ln{\hat{v}_{i}}\bigr]
                           +\bigl[(\ln{\bar{v}_{i+1}})^2-(\ln{\bar{v}_{i}})^2\bigr]\nn\\
         && \;\;\;\;\;\;\; -\bigl[\ln{\bar{v}_{i+1}}\ln{\hat{v}_{i+1}}-\ln{\hat{v}_{i}}\ln{\bar{v}_{i}}\bigr]
                           -\bigl[\ln{\hat{\til{v}}_{i+1}}\ln{\hat{\bar{v}}_{i+1}}-\ln{\hat{\bar{v}}_{i}}\ln{\hat{\til{v}}_{i}}\bigr]\biggr\}.
\eea

On summation, all of these terms will cancel out, leaving zero.

\section{Further remarks}\label{Section_conclusion}
Multi-component quad equations generalizing the one derived in \cite{NijPapCapQui1992}, as well as the one given here, are of course of interest.
The multidimensional consistency was used as integrability criterion to obtain generalizations of the lattice Boussinesq equation in \cite{Hie2011}, and identification of these systems with a generalized dispersion relation within the direct linearization framework \cite{ZhaZhaNij2011} opens up a potentially fruitful avenue in this direction.
As indicated in Section \ref{Section_Symmetry}, the transformation theory of such systems is quite involved even before non-local B\"acklund-type transformations are considered.

There is also a hierarchy presented in \cite{Atk2008} with the lattice Schwarzian KdV and lattice Schwarzian Boussinesq equations as the lowest two members; this system was obtained directly in relation to the modified hierarchy (\ref{discreteMGD}) by generalizing a B\"acklund transformation known in the case $N=1$ since \cite{NijRamGraOht2001}:
\bse
\bea
 \til{z}_{n}-z_{n} & = & \frac{1}{p}\frac{v_{n+1}\til{v}_{n-1}}{v_{n}\til{v}_{n}},\\
 \hat{z}_{n}-z_{n} & = & \frac{1}{q}\frac{v_{n+1}\hat{v}_{n-1}}{v_{n}\hat{v}_{n}},
\eea
\ese
where $n\in\{1,\dots,N\}$ and by our convention $v_0=v_{N+1}=1$ as usual. It can readily be seen that this system is compatible in the $z_{n}$ provided the $v_{n}$ satisfy \eqref{discreteMGD}, and similarly it is compatible in the $v_{n}$ provided the $z_{n}$ satisfy the system
\bse
\bea
 \frac{\hat{z}_{i+1}-z_{i+1}}{\til{z}_{i+1}-z_{i+1}} & = & \frac{\hat{\til{z}}_{i}-\til{z}_{i}}{\hat{\til{z}}_{i}-\hat{z}_{i}}, \;\;\;\;\;\; i\in\{1,\dots,N-1\},\\
 \frac{p^{N+1}}{q^{N+1}}\prod_{i=1}^{N}\frac{\til{z}_{i}-z_{i}}{\hat{z}_{i}-z_{i}} & = & \frac{\hat{\til{z}}_{N}-\til{z}_{N}}{\hat{\til{z}}_{N}-\hat{z}_{N}}.
\eea
\ese
Unlike for the modified hierarchy from which it is derived, the expected Lagrangian formulation of this multidimensionally consistent Schwarzian 
hierarchy is presently not known.

The Lagrangian formulation of the multicomponent quad equation in this paper contributes to the mounting evidence that the closure relation and the  associated least-action principle are key properties of discrete integrable systems (and indeed of the corresponding continuous systems as well, cf. \cite{XenNijLob2010}). An important motivating factor to study such a structure is its potential physical significance; in particular the Lagrangian 
multi-form structure constitutes a possible departure point for the quantization of such integrable models along the line of a path integral formulation.

\section*{Acknowledgments}
JA was supported by the Australian Research Council (ARC) Discovery Grant \#DP110104151. SBL was supported by the ARC Discovery Grant \#DP110100077. This grant also partially supported FWN during a recent visit to 
La Trobe University (Melbourne) where this work was undertaken. He currently holds a Royal Society/Leverhulme Trust Senior Research Fellowship (2011/12).

\end{document}